\documentclass[twocolumn,showpacs,amsmath,amssymb,aps,prb,superscriptaddress]{revtex4-1}

\usepackage[]{cleveref}
\usepackage[]{graphicx}
\usepackage[]{verbatim}
\usepackage[]{color}
\usepackage{xcolor,colortbl}
\usepackage[abs]{overpic}
\usepackage{rotating}
\usepackage{tabularx}

\usepackage{amsfonts}
\usepackage{amsmath}
\usepackage{amssymb}
\usepackage{bm}
\usepackage{graphicx}
\usepackage[breaklinks,colorlinks=true,citecolor=blue]{hyperref}
\usepackage{mathrsfs}
\usepackage[lofdepth,lotdepth,caption=false]{subfig}
\usepackage{varwidth}
\usepackage{wrapfig}
\usepackage{times}
\usepackage{longtable}
\usepackage{multirow}
\usepackage{tikz}

\definecolor{darkblue}{RGB}{0 60 120}
\definecolor{eggplant}{RGB}{190 10 150}
\definecolor{darkgray}{RGB}{70 70 70}
\definecolor{lightgray}{RGB}{200 200 200}
\definecolor{lightgray2}{RGB}{245 215 110}
\definecolor{lightgray3}{RGB}{255 0 0}
\definecolor{blue(pigment)}{rgb}{0.2, 0.2, 0.6}

\graphicspath{{./}}

\begin{document}

\title{Revealing frustrated local moment model for pressurized hyperhoneycomb iridate:\\
paving a way toward quantum spin liquid}

\author{Heung-Sik Kim}
\affiliation{Department of Physics and Center for Quantum Materials , University of Toronto, 60 St.~George St., Toronto, Ontario, M5S 1A7, Canada}

\author{Yong Baek Kim}
\email{ybkim@physics.utoronto.ca}
\affiliation{Department of Physics and Center for Quantum Materials , University of Toronto, 60 St.~George St., Toronto, Ontario, M5S 1A7, Canada}
\affiliation{Canadian Institute for Advanced Research / Quantum Materials Program, Toronto, Ontario MSG 1Z8, Canada}

\author{Hae-Young Kee}
\email{hykee@physics.utoronto.ca}
\affiliation{Department of Physics and Center for Quantum Materials , University of Toronto, 60 St.~George St., Toronto, Ontario, M5S 1A7, Canada}
\affiliation{Canadian Institute for Advanced Research / Quantum Materials Program, Toronto, Ontario MSG 1Z8, Canada}

\begin{abstract}
There have been tremendous experimental and theoretical efforts toward discovery of quantum spin liquid phase
in honeycomb-based-lattice materials with strong spin-orbit coupling. Here the bond-dependent Kitaev interaction
between local moments provides strong magnetic frustration and if it is the only interaction present in the system, 
it will lead to an exactly solvable quantum spin liquid ground state. 
In all of these materials, however, the ground state is in a magnetically ordered phase due to additional 
interactions between local moments. Recently, it has been reported that the magnetic order in hyperhoneycomb
material, $\beta$-Li$_2$IrO$_3$, is 
suppressed upon applying hydrostatic pressure and the resulting state becomes a quantum paramagnet or
possibly a quantum spin liquid. Using {\it ab-initio} computations and strong coupling expansion, we investigate 
the lattice structure and resulting local moment model in pressurized $\beta$-Li$_2$IrO$_3$.
Remarkably, the dominant interaction under high pressure is not the Kitaev interaction nor further neighbor interactions, 
but a different kind of bond-dependent interaction. This leads to strong magnetic frustration and may provide a platform for
discovery of a new kind of quantum spin liquid ground state.
\end{abstract}

\maketitle

\section{Introduction}
Magnetic frustration is often regarded as a prominent route to realize quantum spin liquid states, novel
quantum paramagnetic states with fractionalized excitations\cite{BalentsQSL2010}. In the Kitaev model on the honeycomb lattice,
magnetic frustration is achieved by bond-dependent Ising interactions, where there exist macroscopic
number of classically degenerate ground states\cite{kitaev2006anyons}. The quantum ground state can be solved exactly and 
is shown to be a quantum spin liquid. Recently, much effort has been put forward to 
realize the Kitaev interaction in honeycomb-based-lattice materials with strong spin-orbit coupling\cite{William-review,Jeff-review,
Robert2015}, where the spin-orbit coupling and edge-sharing octahedra structure allow such interactions\cite{jackeli2009mott}.
This physics has been explored in two dimensional honeycomb lattice systems such as 
Na$_2$IrO$_3$\cite{choi2012spin,Ye2012dr,singh2012relevance,mazin_prl,
Chaloupka2013,Yamaji2014,Katukuri2014,Chun2015,Rousochatzakis2015}, 
$\alpha$-Li$_2$IrO$_3$\cite{singh2012relevance,Knolle2014}, RuCl$_3$\cite{plumb2014alpha,Banerjee2016,Winter2016}
as well as three-dimensional hyperhoneycomb $\beta$-Li$_2$IrO$_3$\cite{Mandal2009,Eric1,SBLee2014,
Takayama2014uf,Biffin2014ky,RobertPRL2015,Kim-EPL2015,Katukuri2016,OBrien2016} 
and stripy honeycomb $\gamma$-Li$_2$IrO$_3$\cite{Modic2014ch,kimch2013td,Eric3,Kimchi2015,Perreault2015,Eric4} 
systems. Here the local moments on Ir (or Ru) ions can be described by the pseudospin $j_{\rm eff}$=1/2 
degree of freedom, a spin-orbit entangled Kramers doublet\cite{kim2008novel,kim2009phase}. 

These materials, however, develop magnetic ordering at low temperatures, defying attempts to
achieve quantum spin liquid ground states\cite{choi2012spin,Banerjee2016,Biffin2014ky,Biffin2014_2}. 
It has been shown that such magnetic ordering occurs
due to the presence of other interactions between $j_{\rm eff}$=1/2 moments\cite{rau2014generic,Eric3}. On the other hand, the
nature of the observed magnetic order is strongly dependent on the Kitaev interaction, which is an 
indirect evidence that strength of the Kitaev interaction in these materials is significant.
This suggests that if there is a way to control relative strength of these interactions, 
one may be able to achieve a quantum spin liquid ground state. 

Very recently, hydrostatic pressure was applied to the hyperhoneycomb material, $\beta$-Li$_2$IrO$_3$,
and it was found that the magnetic order disappears for sufficiently high pressure while the material 
remains insulating\cite{Takagi-talk}.
The NMR and specific heat measurements found no signature of any broken symmetry, 
which could be regarded as a sign of a possible quantum spin liquid ground state. 
Hence the question is what kind of local moment interactions are present in the high pressure phase
and whether such interactions would lead to a quantum spin liquid ground state. 

In this article, we theoretically investigate the lattice structure and local moment model for $\beta$-Li$_2$IrO$_3$
under hydrostatic pressure using {\it ab-initio} density functional theory (DFT) computations and strong coupling expansion.
It is shown that the dominant
interaction between local moments in high pressure structure is the so-called symmetric anisotropic (SA) interaction 
which depends on bond-directions, as explained below. On the other hand, the usual Heisenberg and Kitaev interactions are 
generally suppressed and, in contrast to a naive expectation, further neighbor interactions are not so significant. 
If only the SA interaction is present, the classical version of the model is highly frustrated and there exists
macroscopic degeneracy of classically degenerate ground states. Interestingly, the manifold of classically degenerate
states in the SA model is very different from that of the Kitaev model\cite{Eric3}. This points to an interesting possibility that
the quantum version of such a model may support a quantum spin liquid state that is distinct from the Kitaev 
spin liquid state.
 
\begin{figure}
\centering
\includegraphics[width=0.4 \textwidth]{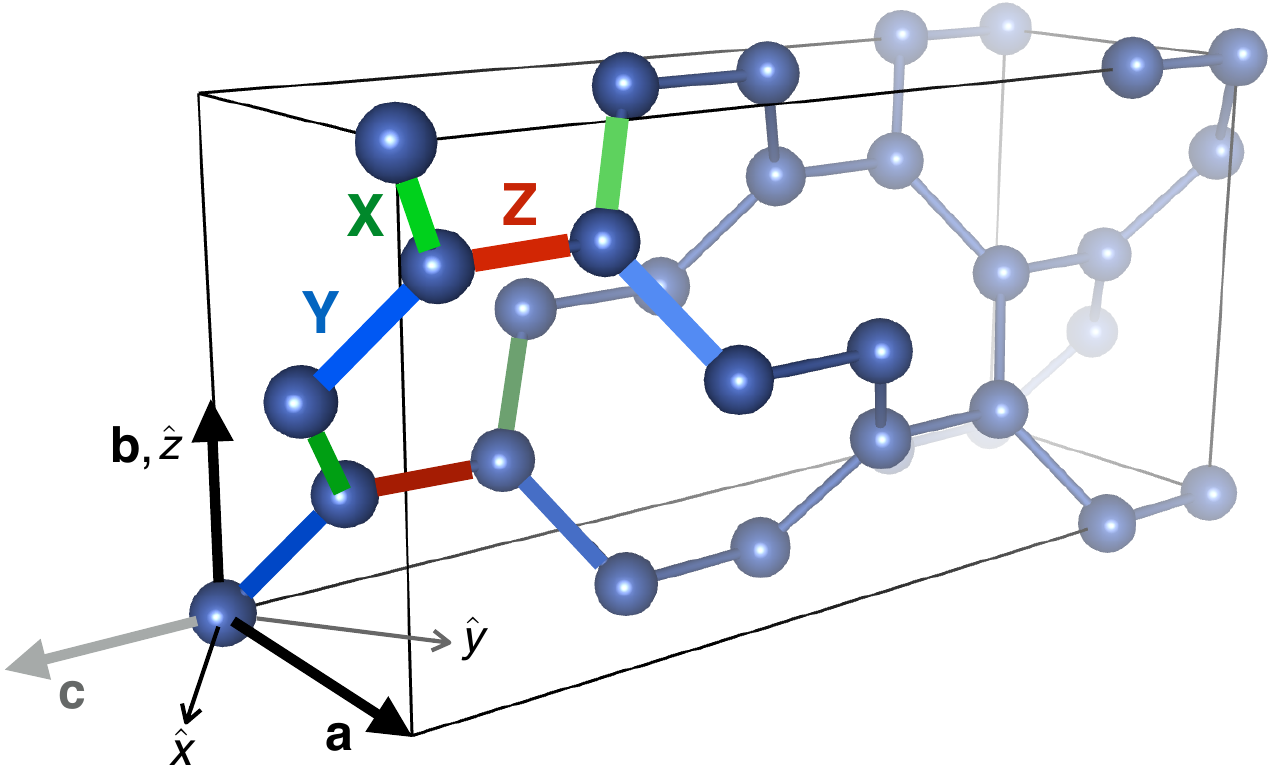}
\caption{(Color online) 
Hyperhoneycomb network of Ir atoms in $\beta$-Li$_2$IrO$_3$ shown in the conventional 
orthorhombic unit cell (depicted in black lines). 
Each site is connected to three nearest-neighbor sites by X (green), Y (blue), and Z (red)-type bonds.
Note that, X- and Y-bonds are equivalent under $C_2$ rotations. ${\bf a}$ and 
${\bf b}$ lattice vectors, which are more sensitive than {\bf c} lattice vector to pressure, 
are depicted as black arrows. $\hat{x}$,  $\hat{y}$, 
and  $\hat{z}$ are local cubic axes.
}
\label{fig:str}
\end{figure}

More specifically, we find that the space group of the {\it ab-initio} optimized lattice structure
remains unchanged ({\it Fddd}, SG. 70) under pressure at least up to 10.2 GPa 
while the lattice parameters become more anisotropic compared to those at ambient
pressure. The local moment model in the strong coupling limit has the following general form\cite{rau2014generic},
\begin{equation} \label{eq:Hspin}
H = \sum_{\langle ij \rangle \in \alpha \beta (\gamma)} 
\left [ J_{ij} {\bf S}_i \cdot {\bf S}_j + K_{ij} S_i^{\gamma} S_j^{\gamma} \pm 
\Gamma_{ij} (S_i^{\alpha} S_j^{\beta} + S_i^{\beta} S_j^{\alpha}) \right ],
\end{equation}
where ${\bf S}_i$ is the $j_{\rm eff}$ = 1/2 pseudospin at site $i$, the summation
is over the nearst-neighbor (NN) bonds $\langle i, j \rangle$ labelled by $\gamma \in ({\rm X, Y, Z})$, and 
$\langle i, j \rangle \in \alpha \beta (\gamma)$ is shorthand for 
$\langle i, j \rangle \in \gamma, \alpha \not= \beta \not= \gamma$. 
The $\pm$ sign in front of $\Gamma$ is a reminder that, unlike the $J$ and $K$ terms, the
$\Gamma$ term can have relative minus signs on different bonds (the
sign structures of $\Gamma$ are explained in Ref. \onlinecite{Eric3}).
Here $J$, $K$, and $\Gamma$ represent the Heisenberg, Kitaev, and SA interactions,
respectively. At ambient pressure, 
the magnitudes of $J$, $K$, and $\Gamma$ are uniform for all of X, Y, and Z-type bonds (see Fig. \ref{fig:str}
for NN bonds and Fig. \ref{fig:JKG} for the magnitudes). 
When $K$ is the dominant interaction with non-zero
$\Gamma$ and $J$, this model can explain the incommensurate counter-rotating spiral order
observed in a resonant elastic X-ray scattering experiment\cite{Eric3}.
A previous DFT computation shows that the material in the experimentally determined
structure is indeed in this parameter regime\cite{Kim-EPL2015}.

Upon increasing pressure, DFT results indicate that the bond lengths of the X, Y-type bonds become
shorter than that of the Z-type bond. The biggest change occurs in $t_{dd\sigma}$-type
hopping integral, which represents a direct overlap in the $\sigma$ bonding channel 
between $d$ orbitals at NN sites. As explained later, this change 
makes the Kitaev and Heisenberg interactions much smaller and these interactions on
X, Y and Z-type bonds become anisotropic. The dominant interaction, however,
is the SA interaction $\Gamma$ while it becomes also bond-anisotropic.
In addition, further neighbor interactions are found to be, in general, 
less than 10\% of the NN interactions. Hence it is clear that a good starting
point for the local moment model at high pressure is the SA interaction model, which is
highly frustrated at the classical level and holds a promise for a quantum spin liquid state.

\section{\label{app:dft_details}Computational details}
For the electronic structure calculations, we employ 
the Vienna {\it ab-initio} Simulation Package ({\sc vasp}), which uses the 
projector-augmented wave (PAW) basis set\cite{VASP1,VASP2}. 
520 eV of plane wave energy cutoff is used, and for
$k$-point sampling 9$\times$9$\times$9 grid including Gamma point
is employed for the primitive cell. On-site Coulomb interaction is
incorporated using the Dudarev's rotationally invariant DFT+$U$ formalism\cite{Dudarev}
with effective $U_{\rm eff} \equiv U-J = 2$ eV. We employ two different trial
magnetic configurations; N\'{e}el-type and zigzag-type antiferromagnetic orders, which yield
the same result. For each configuration
with different cell volume and magnetism, structural optimization for the cell shape and internal coordinates is
performed with a force criterion of 1 meV / \AA~and without any symmetry constraints. 
A revised Perdew-Burke-Ernzerhof generalized gradient 
approximation (PBEsol)\cite{PBEsol} is used for structural optimizations and total energy 
calculations, which yields the best agreement of calculated lattice parameters to the 
experimental ones in conjunction with SOC and $U_{\rm eff}$\cite{Takayama2014uf}. 
Optimized structures are tabulated in Table \ref{tabA:str} in Appendix \ref{app:structure}.
After the structural optimizations, the hopping integrals between the Ir $t_{\rm 2g}$ orbitals
are computed by employing maximally-localized Wannier orbital 
(MLWF) formalism\cite{MLWF1,MLWF2} implemented in Wannier90 package\cite{Wannier90},
but without including $U_{\rm eff}$ and magnetism. 
The computed $t_{\rm 2g}$ Wannier hopping integrals are presented in Table \ref{tabA:hops}
in Appendix \ref{app:jeffhops}.
{\sc vesta}\cite{VESTA} package was used to draw the crystal structure in Fig. \ref{fig:str}.

It should be mentioned that, structure optimizations in the absence of SOC or the Coulomb interaction
lead to severe Ir-Ir dimerization of the Z-bond, regardless of the choice 
of exchange-correlation functionals and other parameters. Since such dimerization has not 
been observed in experimental crystal structures\cite{Takayama2014uf,Biffin2014ky}, we 
conclude that both SOC and Coulomb interaction are crucial in maintaining
the observed hyperhoneycomb structure in $\beta$-Li$_2$IrO$_3$. Note that, similar suppression of 
dimerization due to SOC was reported in the quasi-two-dimensional $\alpha$-RuCl$_3$, 
which has the similar local geometry of edge-sharing metal-anion octahedra\cite{Kim2016}.

\begin{figure}
\centering
\includegraphics[width=0.35 \textwidth]{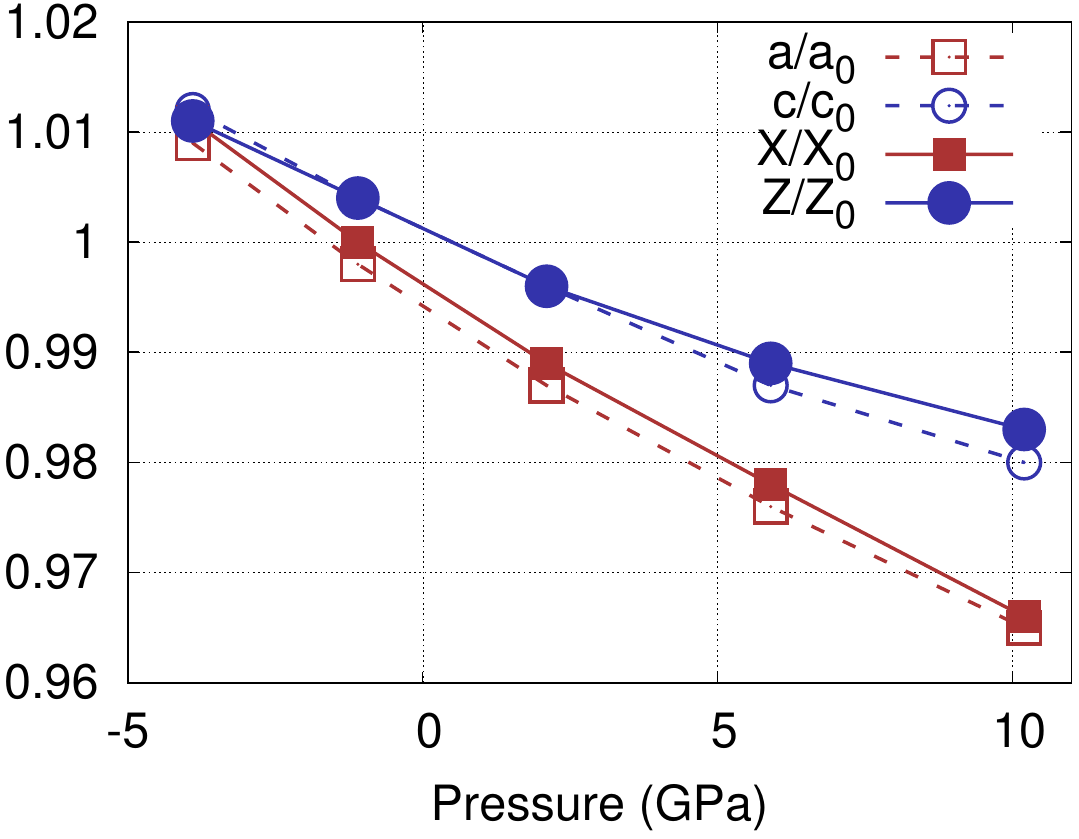}
\caption{(Color online) 
Pressure-dependence of the ratios of lattice constants and NN bond lengths
with respect to the experimental values $\{{\bf a}_0, {\bf c}_0, {\rm X}_0, {\rm Z}_0\}$
at ambient pressure, reported in Ref. \onlinecite{Takayama2014uf}. 
Note that ${\bf a}/{\bf a}_0 \simeq {\bf b} / {\bf b}_0$ and $d_{\bf X} = d_{\bf Y}$.  }
\label{fig:acXZ}
\end{figure}

\section{Evolution of crystal structure under pressure}
Fig. \ref{fig:acXZ} shows the evolution of the ${\bf a}$ and ${\bf c}$ lattice constants and NN bond lengths with 
respect to the hydrostatic pressure (more details about optimized crystal structures are in Table \ref{tabA:str}
in Appendix A). 
In Fig. \ref{fig:acXZ}, the volume of the unit cell is reduced from 
103\% to 91\% with a decrement of 3\%, where the
largest and smallest volume correspond to the pressures of P = -3.9 and 10.2 GPa in our calculations. Note that
the ratio between ${\bf a}$ and ${\bf c}$ lattice parameters becomes closer to the
experimental ${\bf a}_0 / {\bf c}_0$ at P = -3.9 GPa, hence we take this pressure as a reference point. 
It is shown in the figure that the ${\bf a}$ 
(and ${\bf b}$) lattice parameters are reduced by $\sim$ 1.5 \% more than the ${\bf c}$ parameter, implying 
the X- and Y-bonds, forming the zigzag chains in the hyperhoneycomb structure (shown in Fig. \ref{fig:str}), are more compressed 
than the Z-bonds. Indeed, the X-bond length, denoted as $d_{X}$, is twicely more compressed than the Z-bond length $d_{Z}$
at P = 10.2 GPa;
$d^0_{\rm X} - d_{X}$ and $d^0_{\rm Z} - d_{Z}$ being 3.4 and 1.7\% of the experimental $d^0_{\rm X}$ and $d^0_{\rm Z}$ at ambient 
pressure, respectively\cite{Takayama2014uf}. 

Compared to the lattice constants and the NN Ir-Ir bond lengths, the Ir-O bond lengths show smaller changes. 
$d^{Z}_{\rm Ir-O}$ and $d^{X}_{\rm Ir-O}$, the Ir-O bond lengths participating in the NN Z- and X-bonds, 
are reduced by $\sim$ 1.4\% and 1.2\% respectively when P is increased from -3.9 to 10.2 GPa. These changes are smaller
compared to the $\sim$ 3 to 4.5 \% reduction of the NN Ir-Ir bond lengths,
From this comparison, it can be deduced that the direct hopping channels due to the direct overlap of neighboring Ir
$t_{\rm 2g}$ orbitals, which are relevant to the Ir-Ir bond length, should be more affected by hydrostatic pressure 
than the oxygen-mediated indirect channels, relevant to the Ir-O bond length. This is confirmed
in the computation of the hopping integrals, as presented in the next section. 


\section{$t_{\rm 2g}$ hopping channels}
The hopping integrals between the NN Ir $t_{\rm 2g}$ orbitals $\{d_{xz}, d_{yz}, d_{xy}\}$ for the X- and Z-bonds, 
represented by $3\times 3$ matrices, are as follows.
\begin{align}
\hat{T}^{\rm Z} &= \left(
\begin{array}{rrr}
t_1 & t_2 & t_i \\
t_2 & t_1 & -t_i \\
-t_i & t_i & t_3 
\end{array}
\right),~
\hat{T}^{\rm X} = \left(
\begin{array}{rrr}
t_3 & t_4 & t'_4 \\
t_4 & t_1 & t_2 \\
t'_4 & t_2 & t'_1 
\end{array}
\right), 
\end{align}
where the forms of $\hat{T}^{\rm Z}$ and $\hat{T}^{\rm X}$ are determined by the point group symmetries
at the Z- and X-bond centers\cite{Takayama2014uf,Eric3}. 
Note that, $\hat{T}^{\rm Y}$ can be obtained by applying twofold rotations to $\hat{T}^{\rm X}$. 
Here the most dominant terms are $t_1$ ($t'_1$), 
$t_2$, and $t_3$, which originate from $t_{dd\delta}$-like direct, $t_{dpd\pi}$-like indirect, and $t_{dd\sigma}$-like 
direct overlaps, respectively. The sign of $t_2$ term at the X-bond flips when the twofold rotations along the $\hat{z}$ and
${\bf a} \parallel \hat{x}-\hat{y}$ axes are applied, hence we show only the value of $\vert t_2 \vert$ hereafter. 
Other minor components, $t_i$ and $t_4$ (and $t'_4$) come from trigonal distortions, where the
antisymmetric $t_i$ terms in $\hat{T}^{\rm Z}$ arise due to the absence of inversion at the Z-bond center. 
Detailed illustrations for such terms 
in $\beta$-Li$_2$IrO$_3$ are presented in Ref. \onlinecite{Kim-EPL2015}. Note that, since Ir-O-Ir bond angles become 
closer to 90$^\circ$ when pressure is increased, the magnitudes of $t_i$, $t_4$, and $t'_4$ terms are reduced below 
5\% of that of the largest hopping term. The difference between $t_1$ and $t'_1$ at the X-bond also reduces from 
$\sim$ 20 to 6\% of the average of $t_1$ and $t'_1$ as P is increased from -3.9 to 10.2 GPa. Hence, hereafter we denote $t_1$ 
as the averaged value of $t_1$ and $t'_1$ and present the evolution of $t_{1,2,3}$ as a function of P. 

\begin{figure}
\centering
\includegraphics[width=0.35 \textwidth]{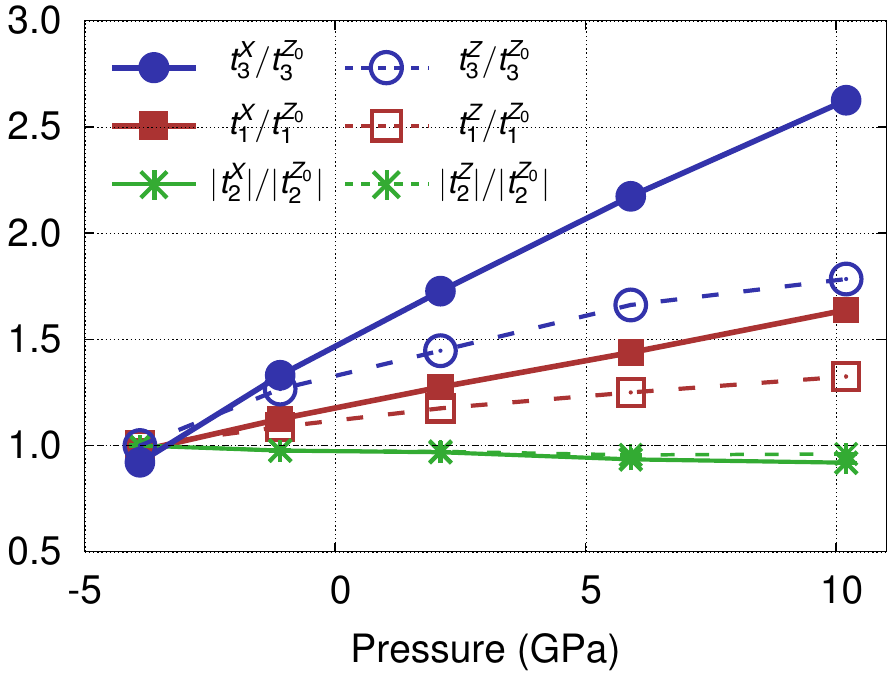}
\caption{(Color online) 
Pressure-dependence of the ratios of three Ir $t_{\rm 2g}$ hopping integrals for the X- and Z-bonds, 
with respect to $t^{\rm Z_0}_{1,2,3}$ denoting the $t_{1,2,3}$ channels for the Z-bond at P = -3.9 GPa, respectively. 
Solid and dashed lines depict the evolution of hopping amplitudes in the X- and Z-bonds, respectively.
Note that, at P = -3.9 GPa, the NN hopping terms are isotropic. }
\label{fig:t123}
\end{figure}

Fig. \ref{fig:t123} shows the evolution of the ratios $t_{1,2,3} / t^{\rm Z_0}_{1,2,3}$ with respect to pressure,
where $t^{\rm Z_0}_{1,2,3}$ are the values of Z-bond hopping terms at P = -3.9 GPa 
($t^{\rm Z_0}_{1} = 80$ meV, $\vert t^{\rm Z_0}_{2} \vert = 248$ meV, and $t^{\rm Z_0}_{3} = -139$ meV).  
As expected in the previous section, the $t_{dd\sigma}$-like $t_3$ channel shows the largest enhancement of 260\%
at the X-bond. Due to the larger
compression of the X-bond compared to the Z-bond, $t^{\rm X}_{3}$ becomes 75\% larger than $t^{\rm Z}_{3}$. 
This huge enhancement makes $t_3$ the dominant hopping term at P = 10.2 GPa;
-365 and -248 meV for the X- and Z-bonds respectively.
The $t_{dd\delta}$-like $t_1$ channel is also increased by the pressure, with smaller enhancement compared to
$t_3$. On the contrary, $t_2$ channel is almost unchanged with the small decrease of 4 $\sim$ 8\% at P = 10.2 GPa, 
due to the cancellation between the $t_{dpd}$-like indirect and $t_{dd}$-like direct overlaps within the $t_2$ channel. 
As suggested in other systems with similar local crystal structure\cite{Kim2016,Winter2016},
these changes in NN hopping channels greatly affect the magnetic exchange interactions between the $j_{\rm eff}$ = 1/2
pseudospins in the strongly correlated regime, as we will discuss in the following section. 

It should be mentioned that, compared to these huge changes in the NN channels, the second, third, and further-neighbor 
channels do not show any significant changes. For example, the largest NN hopping
term ($t^{\rm II}_{\rm NNN}$ in Ref. \onlinecite{Kim-EPL2015}) is enhanced from 77 to 78 meV as P is increased from -3.9 
to 10.2 GPa. The largest third- ($t^{\rm II}_{\rm 3NN}$ in Ref. \onlinecite{Kim-EPL2015}) and fourth-neighbor terms, 
corresponding to  
-45 and -31 meV at P = -3.9 GPa, respectively, are enhanced at most by 15 meV as P is increased.
From these results, we conclude that the role of further-neighbor terms is not significant in the pressure-induced
paramagnetic phase of $\beta$-Li$_2$IrO$_3$. 

\begin{figure}
\centering
\includegraphics[width=0.35 \textwidth]{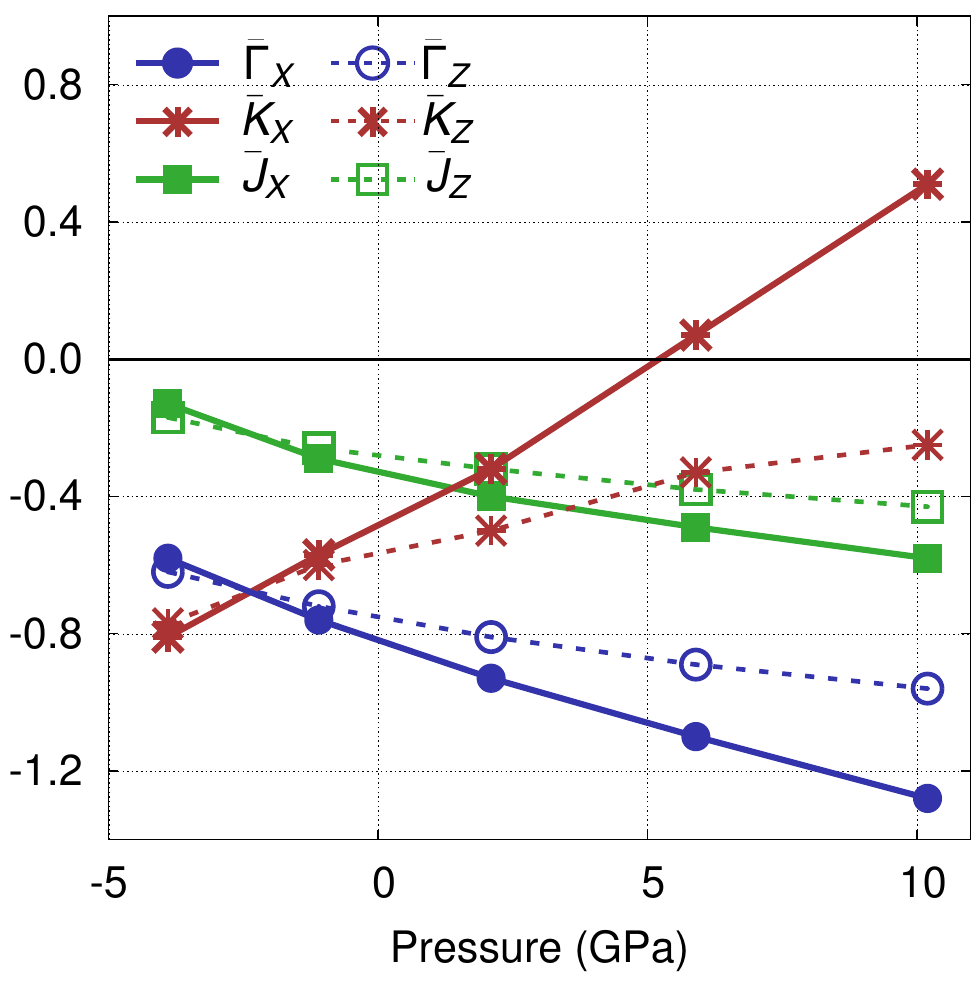}
\caption{(Color online) 
Pressure-dependence of the exchange interactions for the $j_{\rm eff}$ = 1/2 pseudospins. 
Note that dimensionless values 
$\{\bar{J},\bar{K},\bar{\Gamma} \}_{\rm X,Z} \equiv \{ J,K,\Gamma \}_{\rm X,Z} / \sqrt{(J^0_Z)^2 + (K^0_Z)^2 + (\Gamma^0_Z)^2}$
are shown, where $\{J,K,\Gamma \}^0_{\rm Z}$ are the exchange interactions for the Z-bond 
at P = -3.9 GPa. 
}
\label{fig:JKG}
\end{figure}

\section{Magnetic exchange interactions at high pressure}
The huge changes in the NN hopping channels upon pressure affect the $j_{\rm eff}$ = 1/2 NN exchange 
interactions substantially, where each interaction term in the spin model written in Eq.(1) is represented
as follows\cite{rau2014generic,Eric3}.
\begin{align}
J &= \frac{4}{27} \left[ \frac{(4J_H + 3U)(2t_1 + t_3)^2}{U^2} - \frac{16J_H(t_1-t_3)^2}{(2U+3\lambda)^2} \right],  \\
K &= \frac{32J_H}{9}\left[ \frac{(t_1-t_3)^2 - 3t_2^2}{(2U-3\lambda)^2} \right],
\Gamma = \frac{64J_H}{9}\frac{t_2(t_1-t_3)}{(2U+3\lambda)^2}, 
\end{align}
where $U$, $J_H$, and $\lambda$ are the on-site Coulomb interaction, 
Hund's coupling, and Ir $t_{\rm 2g}$-orbital SOC respectively.
In this study we employ $U$ = 2.0 eV, $J_H / U$ = 0.2, and $\lambda$ = 0.45 eV. 
Note that, apart from the overall energy scale,  
the ratios between the exchange interactions are almost insensitive to $J_H/U$ when $J_H/U > 0.05$.
In principle, additional SA term $\Gamma'$ is allowed to exist, which is proportional to 
$t_4$ (and $t'_4$) as discussed in Ref. \onlinecite{rau-trigonal}, and the DM vector parallel to 
the bond direction at the Z-bond is allowed as well. However, their magnitudes become 
insignificant as pressure is increased. 

Fig.~\ref{fig:JKG}(a) shows the calculated values of the exchange interactions, where all the values are divided 
by the absolute magnitude of $\sqrt{(J^0_Z)^2 + (K^0_Z)^2 + (\Gamma^0_Z)^2}$ along the Z-bond at P = -3.9 GPa
and are shown as dimensionless numbers. Two notable features are found; i) the SA term $\Gamma$ is enhanced significantly
by the pressure. At a relatively low pressure of $\sim$ -2.5 GPa, the SA term overcomes $K$ and becomes the largest term.
It becomes even larger under higher pressure; at P = 2.1 GPa, the ratio between the magnitudes of the exchange 
interactions at the X-bond is $\vert J_{\rm X} \vert : \vert K_{\rm X} \vert: \vert \Gamma_{\rm X} \vert = 0.43:0.34:1$,
where all $J,K,\Gamma < 0$. As such, magnetic properties in the pressurized $\beta$-Li$_2$IrO$_3$ would be distinct
from those of the Kitaev-dominated phases at ambient pressure\cite{Eric3}. 
ii) While the X- and Z-bonds share almost the same values of exchange interactions at P = -3.9 GPa, 
the anisotropy between the X- and Z-bonds becomes significant in the high-pressure regime of P $>$ 5 GPa 
with the sign flip of $K$ on the X-bonds. The anisotropy in the NN interactions becomes larger 
than the strength of further-neighbor exchange interaction terms, hence the anisotropy in the NN interactions
would play a more significant role than further-neighbor interactions in the high-pressure phase of 
$\beta$-Li$_2$IrO$_3$
\footnote{We checked that the alternative expression for the exchange interactions, 
presented in Ref. \onlinecite{Winter2016}, yields consistent results with ours shown in Fig. \ref{fig:JKG}
except for small quantitative difference as to slightly enhanced bond anisotropy.}. 


As noticed in previous studies on $\alpha$-\{Li,Na\}$_2$IrO$_3$ (Ref. \onlinecite{Nishimoto2016,Winter2016}),
the magnitude of exchange interactions is sensitive to the local Ir-O geometry, especially to the
ratio between the NN Ir-Ir and Ir-O bond lengths. This ratio is in turn controlled
by the Ir-O-Ir bond angle for a given Ir-Ir distance. Previous studies revealed 
that the FM Kitaev term is suppressed when the Ir-O-Ir bond angle becomes smaller. 
Ref. \onlinecite{Winter2016} found also that the $\Gamma$ term is 
substantially enhanced as the bond angle is reduced. Since the reduced bond angle 
corresponds to the increased Ir-O distance and the reduced $p$-$d$-hopping amplitude,
their results are consistent with our finding of reduced Kitaev and enhanced $\Gamma$ terms
under pressure. 

Such pressure-induced effects on the anisotropic exchange interactions 
would manifest in the anisotropy of the magnetic susceptibility.
For example, high-temperature expansion of Eq. (1) yields the anisotropic Curie-Weiss temperatures, which
satisfy $\theta^{\rm CW}_{\bf a} - \theta^{\rm CW}_{\bf c} \simeq 2\vert \Gamma_Z \vert$
and $\theta^{\rm CW}_{\bf a} + \theta^{\rm CW}_{\bf c} - 2\theta^{\rm CW}_{\bf b} \simeq
2(K_Z - K_X)$, where $\theta^{\rm CW}_{\bf a,b,c}$ are the Curie-Weiss temperatures (multiplied by $k_{\rm B}$) with external
field parallel to the ${\bf a,b,c}$ axes respectively. Hence the change in anisotropic 
exchange interactions and bond-anisotropy of the Kitaev term under pressure can be detected from 
the anisotropy of high-temperature susceptibility data.

\section{Discussion and Outlook}
In iridates with honeycomb or hyperhoneycomb lattices, the strong spin-orbit coupling and edge-sharing 
oxygen octahedra structure conspire to generate the celebrated Kitaev interaction, which provides 
magnetic frustration and exactly solvable quantum spin liquid ground state. Such physics has been one
of the main driving forces for research on quantum spin liquid phases in this class of materials. 
It is in contrast to a more conventional paradigm, where further neighbor exchange interactions are 
used to engineer magnetic frustration in bipartite lattices such as the honeycomb or hyperhoneycomb lattices. 
In this work, we ask the question whether the bond-dependent interaction or further-neighbor interaction
is mainly responsible for the suppression of magnetic order or appearance of correlated quantum paramagnetic
state under high pressure as discovered in a recent experiment on pressurized $\beta$-Li$_2$IrO$_3$\cite{Takagi-talk}. 
Remarkably, our analyses of {\it ab initio} computations with structure optimizations and strong coupling expansion,
strongly suggest that the bond-dependent
symmetric anisotropic interaction, which is {\it distinct} from the Kitaev interaction, is the dominant player in the magnetic frustration.
Previous studies of the SA interaction on the honeycomb and hyperhoneycomb lattices have shown that
there exists a macroscopically degenerate manifold of classical ground states\cite{rau2014generic,Eric3}. Hence it is 
conceivable that the quantum version of this model may support the emergence of a new kind of quantum spin liquid
ground state. This would be an excellent topic for future studies.

\begin{acknowledgments}
HSK thanks Jeffrey G. Rau for valuable comments. 
This work was supported by the NSERC of
Canada and the Center for Quantum Materials at the University of
Toronto. This work was performed in part (YBK and HYK) at the Aspen Center for Physics,
which is supported by National Science Foundation grant PHY-1066293.
Computations were mainly performed on the GPC supercomputer
at the SciNet HPC Consortium. SciNet is funded by: the Canada
Foundation for Innovation under the auspices of Compute Canada; the
Government of Ontario; Ontario Research Fund - Research Excellence;
and the University of Toronto. 
\end{acknowledgments}

\appendix

\section{\label{app:structure}Optimized crystal structures under the pressure}
As mentioned above, structure optimizations are carried out in the primitive unit cell 
without enforcing any symmetry constraints. However, the optimized structures show practically 
no deviation from the original 
$Fddd$ space group symmetry. The angles between the orthorhombic Bravais lattice vectors in the optimized 
structures do not deviate from the right angle ($\vert \delta \theta \vert < 0.0004^\circ$). 
{\sc findsym} package\cite{Findsym} is employed for refining the optimized structures, 
and the difference of internal coordinates between structures before and after the refinement 
is smaller than 0.0002 \AA~ for each site. Therefore we conclude that the optimized 
structure under pressure remains in $Fddd$ symmetry without any symmetry lowering. 
The refined structures are presented in Table \ref{tabA:str}.  

Pressure dependence of ${\bf b}$ needs a comment; In Fig. 2 in the main text, only the pressure dependence of 
${\bf a}$ is presented. Compression of ${\bf b}$ is similar to that of ${\bf a}$, where ${\bf a}/{\bf a}_0$ and 
 ${\bf b}/{\bf b}_0$ are 0.968 and 0.962, respectively, at P = 10.2 GPa. ${\bf b}$ is slightly more compressed 
 than ${\bf a}$, but since the compression of ${\bf a}$ and ${\bf b}$ is similar and significantly larger than 
 that of ${\bf c}$, we present ${\bf a}$ as the representative.

\renewcommand*{\arraystretch}{1.4}
\begin{table}[htb!]
\centering
\begin{tabular}{llrrrrr}  \hline\hline
& P (GPa) &~~~~~ -3.9 &~~~~~ -1.1 &~~~~~ 2.1 &~~~~~ 5.9 &~~~~~ 10.2 \\[-3pt]
& $V/V_0$ & 1.03 & 1.00 & 0.97 & 0.94 & 0.91 \\ \hline
                & {\bf a}  & 5.964   & 5.908    & 5.848   & 5.790   & 5.729 \\[-3pt]
                & {\bf b} & 8.545   & 8.440    & 8.340   & 8.238   & 8.137 \\[-3pt]
                 & {\bf c} (\AA) & 18.037 & 17.891 & 17.747 & 17.603 & 17.463 \\[3pt]
Ir   ($16g$)  & $z$ & 0.7085 & 0.7085 & 0.7086 & 0.7087 & 0.7088 \\[3pt]
Li1 ($16g$)  & $z$ & 0.0441 & 0.0448 & 0.0454 & 0.0458 & 0.0460 \\[3pt]
Li2 ($16g$)  & $z$ & 0.8769 & 0.8775 & 0.8779 & 0.8781 & 0.8781 \\ [3pt]
O1 ($16e$)  & $x$ & 0.8561 & 0.8588 & 0.8614 & 0.8637 & 0.8658 \\[3pt]
O2 ($32h$)  & $x$ & 0.6335 & 0.6320 & 0.6305 & 0.6289 & 0.6271 \\[-3pt]
                      & $y$ & 0.3631 & 0.3654 & 0.3676 & 0.3698 & 0.3719 \\[-3pt]
                      & $z$ & 0.0378 & 0.0384 & 0.0390 & 0.0397 & 0.0403 \\ \hline
$d_{\rm Ir-Ir}$ & Z & 3.011 & 2.988 & 2.967 & 2.946 & 2.928 \\[-3pt]
(in \AA)           & X & 3.005 & 2.973 & 2.940 & 2.907 & 2.874 \\[3pt]
$d_{\rm Ir-O}$ & Z & 2.041 & 2.035 & 2.028 & 2.020 & 2.012 \\[-3pt]
(averaged)      & X & 2.034 & 2.029 & 2.023 & 2.017 & 2.010 \\[3pt]
$\theta_{\rm Ir-O-Ir}$ & Z & 95.06 & 94.50 & 94.04 & 93.64 & 93.39 \\[-3pt]
(degree)                     & X & 95.23 & 94.23 & 93.25 & 92.26 & 91.24 \\ \hline\hline
\end{tabular}
\caption{Table of optimized lattice parameters and internal coordinates of pressurized $\beta$-Li$_2$IrO$_3$
with $Fddd$ (SG. 70, origin choice 2) space group symmetry,
where the internal coordinates for each inequivalent site are $(1/8, 1/8, z)$ for Ir and Li1/2, $(x, 1/8, 1/8)$ for O1, 
and $(x, y, z)$ for O2. $V$ and $V_0$ denote the cell volume for the optimized structure at the given pressure
and that of experimental one at the ambient pressure, respectively.
In addition, Ir-Ir and Ir-O bond lengths and Ir-O-Ir bond angles in each NN bond are shown below.
}
\label{tabA:str}
\end{table}

\section{\label{app:jeffhops}$t_{\rm 2g}$ hopping integrals}

The Ir $t_{\rm 2g}$ hopping integrals for two structures at P = -3.9 and 10.2 GPa are 
shown in Table \ref{tabA:hops}. The values are calculated without including $U_{\rm eff}$
and magnetism for each optimized structure. 

\renewcommand*{\arraystretch}{1.4}
\begin{table}
\centering
\begin{tabular}{lrrrrrrr}  \hline\hline
Kind  &~~~~~~~~~~& \multicolumn{3}{r}{P = -3.9 GPa}  &   \multicolumn{3}{r}{P = 10.2 GPa} \\ 
                                &&  \multicolumn{3}{r}{$(V = 1.03V_0)$} & \multicolumn{3}{r}{$(V=0.91V_0)$} \\ [5pt]
\hline
$t_{\rm NN}$, X &&                  ~~~                   $d_{xz}$ & $d_{yz}$ & $d_{xy}$ & ~~~~~~~~ $d_{xz}$ & $d_{yz}$ & $d_{xy}$ \\
${\bf r}_{ij}$=(-$d$,~0,+$d$),      & $d_{xz}$ & -141 & +21  & +26  &                                              -377     &                  & +19 \\  
{Sublat. 1 $\rightarrow$ 4}
&                               $d_{yz}$& +21   & +64  & +249 &                                                         &  +123        & +227 \\  
&                               $d_{xy}$& +26   & +249 & +94 &                                               +19    &  +227        & +140 \\ [2pt]  \arrayrulecolor{lightgray}\hline
$t_{\rm NN}$, Z&&                 ~~~                   $d_{xz}$ & $d_{yz}$ & $d_{xy}$ & ~~~~~ $d_{xz}$ & $d_{yz}$ & $d_{xy}$ \\
${\bf r}_{ij}$=(+$d$,+$d$,~0),  & $d_{xz}$ & +80   & -247 & -25 &                                               +107    & -233          & -11 \\  
{Sublat. 1 $\rightarrow$ 2}
&                             $d_{yz}$& -247  & +80  & +25 &                                              -233     & +107         & +11 \\  
&                             $d_{xy}$& +25   &  -25  & -154 &                                              +12     &  -12           & -262 \\  [2pt]  \arrayrulecolor{darkgray}\hline
$t^{\rm I}_{\rm NNN}$& & ~~~                  $d_{xz}$ & $d_{yz}$ & $d_{xy}$ & ~~~~~ $d_{xz}$ & $d_{yz}$ & $d_{xy}$ \\
${\bf r}_{ij}$=(+$d$,+$2d$,-$d$), & $d_{xz}$ &        &          & -12  &                                                           &                   & \\  
{Sublat. 1 $\rightarrow$ 3}
&                                  $d_{yz}$&         &          & +39 &                                                           &                   & +42 \\  
&                                  $d_{xy}$&  +14 & +62  &        &                                              +16       & +68           & \\  [2pt]  \arrayrulecolor{lightgray}\hline
$t^{\rm II}_{\rm NNN}$& & ~~~                 $d_{xz}$ & $d_{yz}$ & $d_{xy}$ & ~~~~~ $d_{xz}$ & $d_{yz}$ & $d_{xy}$ \\
${\bf r}_{ij}$=(-$d$,+$d$,+$2d$), & $d_{xz}$ &        & +77   &       &                                                            &  +78          & \\  
{Sublat. 1 $\rightarrow$ 1}
&                                   $d_{yz}$& +42 &          & -14 &                                               +43       &                  & \\  
&                                   $d_{xy}$& +14 &          &       &                                                             &                  & \\  [2pt]  \arrayrulecolor{lightgray}\hline
$t^{\rm III}_{\rm NNN}$ & & ~~~               $d_{xz}$ & $d_{yz}$ & $d_{xy}$ & ~~~~~ $d_{xz}$ & $d_{yz}$ & $d_{xy}$ \\
${\bf r}_{ij}$=(-$d$,+$d$,-$2d$), & $d_{xz}$ &        & +24  & +11 &                                               +11      &      +33     & +15 \\  
{Sublat. 1 $\rightarrow$ 1}
&                                  $d_{yz}$& +32&         &        &                                               +46      &      +11      & \\  
&                                  $d_{xy}$&        & -11   &       &                                                             &      -15      & \\  [2pt]  \arrayrulecolor{lightgray}\hline
$t^{\rm IV}_{\rm NNN}$ & & ~~~               $d_{xz}$ & $d_{yz}$ & $d_{xy}$ & ~~~~~ $d_{xz}$ & $d_{yz}$ & $d_{xy}$ \\
${\bf r}_{ij}$=(+$d$,-$2d$,+$d$), & $d_{xz}$ &      &         &        &                                                             &                 & +11 \\  
{Sublat. 1 $\rightarrow$ 4}
&                                  $d_{yz}$&      &         & -24 &                                                             &                 & -34 \\  
&                                  $d_{xy}$&      & -24  & +13  &                                                +11     &  -34         & +16 \\  [2pt]  \arrayrulecolor{darkgray}\hline
$t^{\rm I}_{\rm 3NN}$ & & ~~~                  $d_{xz}$ & $d_{yz}$ & $d_{xy}$ & ~~~~~ $d_{xz}$ & $d_{yz}$ & $d_{xy}$ \\
${\bf r}_{ij}$=(~0,+$2d$,-$2d$),    & $d_{xz}$ &     & -15  & -12 &                                                              &   -14        & -13 \\  
{Sublat. 1 $\rightarrow$ 2}
&                                    $d_{yz}$&     & -36  & -15 &                                                               &  -50        & -16 \\  
&                                    $d_{xy}$&     & -13 &        &                                                -10          & -14         & \\  [2pt]  \arrayrulecolor{lightgray}\hline
$t^{\rm II}_{\rm 3NN}$ & & ~~~                 $d_{xz}$ & $d_{yz}$ & $d_{xy}$ & ~~~~~ $d_{xz}$ & $d_{yz}$ & $d_{xy}$ \\
${\bf r}_{ij}$=(-$2d$,+$2d$,~0), &    $d_{xz}$ &       & -13 &           &                                                         &   -16        & \\  
{Sublat. 1 $\rightarrow$ 1}
&                                   $d_{yz}$&        &        & -11    &                                                        &                 & \\  
&                                   $d_{xy}$& +11 &       & -46   &                                                         &                 & -60 \\ 
\arrayrulecolor{black}\hline\hline
\end{tabular}
\caption{A subset of Ir $t_{\rm 2g}$ hopping terms $\hat{T}_{ij}$ as representatives of each hopping channel up to third NN,
where $\mathcal{H}_{\rm hop} = \sum_{ij} {C}^{\dag}_{i} \cdot \hat{T}_{ij} \cdot {C}_j$
and $\{{C}^{\dag},{C}\}$ being the creation and annihilation operators for $t_{\rm 2g}$ states, respectively.
$d$ is approximate distance between Ir and O. Other hopping terms can be recovered by applying $\hat{T}_{ji} = \hat{T}^{\dag}_{ij}$,
$C^{\bf a,b,c}_{2}$ rotations, and inversion. Values are in meV unit, and terms smaller than 0.5 meV are not shown.}
\label{tabA:hops}
\end{table}

\bibliography{bLIO}

\end{document}